\documentclass[amsmath,amssymb,latexsym,bbm]{llncs}
\usepackage{xypic,epsf,amsmath,amssymb,latexsym,bbm,multicol}
\usepackage[english]{babel}
\usepackage{graphics}
\usepackage{latexsym}
\usepackage{amsmath}
\usepackage{amsfonts}
\usepackage{amssymb}
\usepackage{enumerate}
\newtheorem{thm}{Theorem}
\newtheorem{defi}[thm]{Definition}
\newtheorem{lem}[thm]{Lemma}

\newtheorem{cor}[thm]{Corollary}

\newtheorem{rem}[thm]{Remark}
\newtheorem{ex}[thm]{Example}

\newcommand{\ignore}[1]{}

\newcommand{\Proof}{\noindent {\it Proof\/}:\ }
\newcommand{\QED}{\hspace*{\fill}$\Box$}

\begin{document}
\title{On combinations of local theory extensions} 

\author{Viorica Sofronie-Stokkermans}
\institute{Max-Planck-Institut f{\"u}r Informatik, Stuhlsatzenhausweg 85, 
Saarbr{\"u}cken, Germany\\
e-mail: {\tt sofronie@mpi-sb.mpg.de}} 
\maketitle

\begin{abstract}
In this paper we study possibilities of efficient reasoning in 
combinations of theories over possibly non-disjoint signatures. 
We first present a class of theory extensions (called local 
extensions) in which  
hierarchical reasoning is possible, and give 
several examples from computer science and mathematics in which such 
extensions occur in a natural way.
We then identify situations in which
combinations of local extensions of a theory are again local 
extensions of that theory. 
We thus obtain criteria both for recognizing wider classes of local 
theory extensions, and for modular reasoning in combinations of theories
over non-disjoint signatures. 
\end{abstract}

\section{Introduction}

Many problems in mathematics and computer science
can be reduced to proving the satisfiability of 
conjunctions of literals in a background theory (which 
can be the extension of a base theory 
with additional functions -- e.g., free, monotone, or recursively defined -- 
or a combination of theories).
It is therefore very important to identify situations where 
reasoning in complex theories can be done efficiently and accurately. 
Efficiency can be achieved for instance by: 
\begin{itemize}
\item[(1)] reducing the search space (preferably without loosing completeness);
\item[(2)] modular reasoning, i.e.,  delegating some proof tasks which refer to 
a specific theory to provers specialized in handling formulae of 
that theory.  
\end{itemize}
We are interested in identifying situations when both these goals can be
achieved without loss of completeness. 

\paragraph{Controlling the search space.} 
The quest for identifying theories where the search space can be 
controlled without loss of completeness led  McAllester and Givan to 
define {\em local theories}, that is sets $N$ of Horn clauses 
with the property that for any ground clause $G$, $N \models G$ iff
$G$ can be proved already using those 
instances $N[G]$ of $N$ containing only ground terms occurring in $G$ or in  
$N$. For local theories, validity of ground Horn clauses 
can be checked in polynomial time. 
In \cite{BasinGanzinger96,BasinGanzinger01}, 
Ganzinger and Basin 
defined the more general notion of {\em order locality} and 
showed how to recognize (order-)local theories and how to use these results 
for automated complexity analysis.

Similar ideas also occurred in algebra, where the main interest was to 
identify classes of algebras for which the uniform 
word problem is decidable in 
polynomial time. In \cite{Burris95}, Burris proved that 
if a quasi-variety axiomatized by a set ${\cal K}$ of 
Horn clauses has the property 
that {\em every finite partial algebra which is a partial model of the 
axioms in 
${\cal K}$ can be extended to a total algebra model of ${\cal K}$} 
then the uniform word problem for ${\cal K}$ is decidable in polynomial time.
In \cite{Ganzinger-01-lics}, Ganzinger established a link between 
proof theoretic and semantic concepts for polynomial time 
decidability of uniform word problems. He defined two notions 
of locality for equational Horn theories, and established 
relationships between these notions of locality and 
corresponding semantic conditions, referring to embeddability 
of partial algebras into total algebras.

\paragraph{Modular reasoning.} When reasoning in extensions or combinations 
of theories it is very important to find ways of 
delegating some proof tasks which refer to 
a specific theory to provers specialized in handling formulae of 
that theory.  Of particular interest are situations when reasoning 
can be done: 
\begin{itemize}
\item in a hierarchical way (that is, for reasoning in a theory extension
a prover for the base theory can be used as a black-box), or
\item in a modular way (that is, for reasoning in a combination of theories
reasoning in the component 
theories is ``decoupled'', i.e., the information 
about the component theories is never combined and only formulae 
in the joint signature are exchanged between provers for the components). 
\end{itemize}
One of the first methods for modular reasoning in combinations of 
theories, due to Nelson and Oppen \cite{NelsonOppen1979},  
can be applied for combining decision procedures
of \emph{stably infinite} theories over disjoint signatures. 
There were several attempts to extend the completeness results for
modular inference systems for combinations of theories over non-disjoint
signatures. 
In \cite{Ghilardi04-jar} the component theories need to satisfy a
model theoretical compatibility condition with respect to the shared theory.
In \cite{Tinelli03}, similar modularity results are 
achieved if the theories 
share \emph{all} function symbols.
Several modularity results using superposition  
were established for combinations of theories over disjoint signatures in 
\cite{Armando-et-al01,Hillen04,Bonacina-Armando-Schulz05}. 
In \cite{GSW04,GSW-iandc-06} we 
analyzed possibilities of modular reasoning (using special superposition 
calculi) in combination of first-order theories involving both total
and partial functions. 
The calculi are shown to be complete provided that functions 
that are not in the intersection of the component signatures 
are declared as partial.  Cases
where the partial models can always be made total are identified: 
in such cases modular superposition is also complete with respect to the
standard (total function) semantics of the theories. Inspired by 
the link between embeddability and locality established by 
Ganzinger in \cite{Ganzinger-01-lics}, such extensions were called {\em local}.

\paragraph{Reasoning in local theory extensions and their combinations.}
In \cite{GSW04}, \cite{GSW-iandc-06} and, later, 
in \cite{Sofronie-cade-05} we showed that for 
{\em local theory extensions}\/
efficient hierarchic reasoning is possible.
For such extensions the two goals previously mentioned 
can be addressed at the same time: the locality of an extension 
allows to reduce the search space, but at the same time (as a by-product) 
it allows to perform an easy reduction to a proof task in the base theory 
(for this, a specialized prover can be used as a black box). 

Many theories important for computer science or mathematics are local 
extensions of a base theory:  theories of data structures, 
theories of monotone functions or of 
functions satisfying the Lipschitz conditions.
However, often it is necessary to consider complex extensions, with various 
types of functions (such as, for instance, extensions of the theory of real 
numbers with free, monotone and Lipschitz functions). It is important to 
have efficient methods for hierarchic and/or  modular reasoning also for 
such combinations. 
Finding methods for reasoning in combinations of extensions of a base theory 
is far from trivial: as these are usually combinations of theories 
over non-disjoint signatures, classical combination results such as the 
Nelson-Oppen combination method \cite{NelsonOppen1979} 
cannot be applied;  methods for 
reasoning in combinations of theories over non-disjoint signatures 
-- as studied by Ghilardi et al. \cite{Ghilardi04-jar,BaaderGhilardi05} -- 
may also not always be applicable 
(unless the base theory is universal and the extensions 
satisfy certain model-theoretic compatibility conditions required 
in \cite{Ghilardi04-jar,BaaderGhilardi05}). 

\medskip
\noindent 
In this paper we 
identify situations in which
a combination of local extensions of a base theory is guaranteed to be 
itself a local extension of the base theory. 
We thus obtain criteria  for recognizing complex local 
theory extensions, and for efficient reasoning in such 
combinations of theories (over non-disjoint signatures) in a modular way. 

\medskip
\noindent {\em Structure of the paper:}\/ The paper is structured as follows:
Section~\ref{prelim} contains generalities on partial algebras, 
weak validity and embeddability of partial algebras into total algebras. 
In Section~\ref{sect:local-ext}
 the notion of local theory extension is introduced. 
In Section~\ref{embed} links between embeddability and locality of an 
extension are established. In Section~\ref{examples}, 
examples of local theory extensions are given.
In the following two sections we identify situations under which 
a combination of local extensions of a base theory is guaranteed to be 
itself a local extension of the base theory, 
under stronger (Section~\ref{comp}) or 
weaker (Section~\ref{emb}) embeddability conditions for the  
components. Some ideas on hierarchical and modular reasoning in 
such combinations are discussed in 
Section~\ref{hierarchic}. Section~\ref{conclusions} contains 
conclusions and plans for future work.

The results on combinations 
of local extensions of a base theory presented in this paper 
generalize results on combinations of local theories obtained 
in \cite{Sofronie-Ganzinger-unpublished}.

\section{Preliminaries}
\label{prelim}
This section contains the main notions and definitions necessary in the paper.

\subsection{Partial structures}

Let $\Pi = (\Sigma, {\sf Pred})$ be a signature where $\Sigma$ is a set of 
function symbols and ${\sf Pred}$ a set of predicate symbols. 
\begin{defi}
A {\em partial $\Pi$-structure} is a structure 
$(A, \{ f_A \}_{f \in \Sigma}, \{ P_A \}_{P \in {\sf Pred}})$, 
where $A$ is a non-empty set and for every $f \in \Sigma$ with arity $n$, 
$f_A$ is a partial function from $A^n$ to $A$. The structure is a {\em (total)  structure} if all functions $f_A$ are total.
\end{defi}
In what follows we usually denote both an 
algebra and its support with the same symbol.
Details on partial algebras can be found in \cite{Burmeister}.

The notion of evaluating a term $t$ with respect to a variable assignment 
$\beta : X \rightarrow A$ for its variables in a partial algebra $A$ 
is the same as for total algebras, except that this evaluation is undefined
if $t = f(t_1, \dots, t_n)$ and either one of $\beta(t_i)$ is undefined, or 
else $(\beta(t_1), \dots, \beta(t_n))$ is not in the domain of $f_A$.

\begin{defi}
We define {\em weak validity} in structures 
$(A, \{ f_A \}_{f \in \Sigma}, \{ P_A \}_{P \in {\sf Pred}})$, 
where ${\sf Pred}$ is a set of predicate symbols and 
$(A, \{ f_A \}_{f \in \Sigma})$ is a partial $\Sigma$-algebra.
Let $\beta : X \rightarrow A$.
 \begin{itemize}
\item[(1)] $(A, \beta) \models_w t \approx s$ if and only if one of the conditions below is fulfilled: 
\begin{itemize}
\item[(a)] $\beta(t)$ and $\beta(s)$ are both defined and 
equal; or 
\item[(b)] at least one of $\beta(s)$ and $\beta(t)$ is undefined.
\end{itemize}
\item[(2)] $(A, \beta) \models_w t \not\approx s$ if and only if one of the conditions below is fulfilled: 
\begin{itemize}
\item[(a)]  $\beta(t)$ and $\beta(s)$ are both defined and different; or 
\item[(b)] at least one of $\beta(s)$ and $\beta(t)$ is undefined.
\end{itemize}
\item[(3)] $(A, \beta) \models_w P(t_1, \dots, t_n)$ 
if and only if one of the conditions below is fulfilled: 
\begin{itemize}
\item[(a)]  $\beta(t_1), \dots, \beta(t_n)$ are all defined and
$(\beta(t_1), \dots, \beta(t_n)) {\in} P_A$; or 
\item[(b)] at least one of $\beta(t_1)$, $\dots,$ $\beta(t_n)$ is undefined.
\end{itemize}
\item[(4)] $(A, \beta) \models_w \neg P(t_1, \dots, t_n)$ 
if and only if one of the conditions below is fulfilled: 
\begin{itemize}
\item[(a)]  $\beta(t_1),$ $\dots,$ $\beta(t_n)$ are all defined and
$(\beta(t_1), \dots, \beta(t_n)) \not\in P_A$; or 
\item[(b)] at least one of $\beta(t_1), \dots, \beta(t_n)$ is undefined.
\end{itemize}
\end{itemize}
{\em $(A, \beta)$ weakly satisfies a clause  $C$} 
(notation: $(A, \beta) \models_w C$) 
if $(A, \beta) \models_w L$  for at least one literal $L$ in $C$. 
{\em $A$ weakly satisfies $C$} (notation: $A \models_w C$) 
if $(A, \beta) \models_w C$
for all assignments $\beta$.  
{\em $A$ weakly satisfies a set of clauses ${\cal K}$} (notation: $A \models_w {\cal K}$) if 
$A \models_w C$ for all $C \in {\cal K}$.
\end{defi}
\begin{ex}
{\em 
Let $A$ be a partial ${\Sigma}$-algebra, where 
${\Sigma} = \{ {\sf car}/1, {\sf nil}/0 \}$. 
Assume that ${\sf nil}_A$ is defined and 
${\sf car}_A({\sf nil}_A)$ is not defined.
Then 
\ignore{$A \not\models {\sf car}({\sf nil}) \approx {\sf nil}$
(since ${\sf car}_A({\sf nil})$ is undefined in $A$, but 
${\sf nil}$ is defined in $A$); and 
$A \models {\sf car}({\sf nil}) \not\approx {\sf nil}$,} 
$A \models_w {\sf car}({\sf nil}) \approx {\sf nil}$ and 
$A \models_w {\sf car}({\sf nil}) \not\approx {\sf nil}$
(because  one term is not defined in $A$).
}
\end{ex}
\begin{defi}
A {\em weak $\Pi$-embedding} between the partial structures 
$(A, \{ f_A \}_{f \in \Sigma},$ $\{ P_A \}_{P \in {\sf Pred}})$  
and $(B, \{ f_B \}_{f \in \Sigma}, \{ P_B \}_{P \in {\sf Pred}})$
is a total map $i : A \rightarrow B$ such that 
\begin{itemize}
\item whenever 
$f_A(a_1, \dots, a_n)$ is defined then 
$f_B(i(a_1), \dots, i(a_n))$ is defined and 
$i(f_A(a_1, \dots, a_n)) = f_B(i(a_1), \dots, i(a_n))$; 
\item $i$ is injective; 
\item $i$ is an embedding w.r.t.\ ${\sf Pred}$, 
i.e. for every $P \in {\sf Pred}$ 
with arity $n$ and every $a_1, \dots, a_n \in A$,  
$P_A(a_1, \dots, a_n)$ if and only if $P_B(i(a_1), \dots, i(a_n))$. 
\end{itemize}
In this case we say that $A$ weakly embeds into $B$. 
\end{defi}

\subsection{Theories and extensions of theories}

Theories can be regarded as sets of formulae or as sets of models.
Let ${\cal T}$ be a $\Pi$-theory and $\phi, \psi$ be $\Pi$-formulae.
We say that ${\cal T} \wedge \phi \models \psi$ (written also 
$\phi \models_{{\cal T}} \psi$) is $\psi$ is true in all models of 
${\cal T}$ which satisfy $\phi$. 

In what follows we consider extensions of theories, in which the 
signature is extended by new {\em function symbols} (i.e.\ we assume 
that the set of predicate symbols remains unchanged in the extension).
If a theory is regarded as a set of formulae, 
then its extension with a set of formulae is set union. 
If ${\cal T}$ is regarded as a collection 
of models then its extension 
with a set ${\cal K}$ of formulae consists of all structures
(in the extended signature) which are models of ${\cal K}$ and 
whose reduct to the signature of ${\cal T}_0$ 
is in ${\cal T}_0$. 
In this paper we regard theories as sets of formulae. 
All the results of this paper can easily be reformulated 
to a setting 
in which ${\cal T}_0$ is a collection of models.

\smallskip
\noindent 
Let ${\cal T}_0$ be an arbitrary theory with signature 
$\Pi_0 = (\Sigma_0, {\sf Pred})$, 
where the  set of function symbols is $\Sigma_0$. 
We consider extensions
${\cal T}_1$ of ${\cal T}_0$ with signature 
$\Pi = ({\Sigma}, {\sf Pred})$,
where the set of function symbols is $\Sigma = \Sigma_0 \cup \Sigma_1$.
We assume that ${\cal T}_1$ is obtained from ${\cal T}_0$ by 
adding a set ${\cal K}$ of (universally quantified) clauses. 

\begin{defi}[Weak partial model] 
\ignore{A {\em partial model} of ${\cal T}_1$ with totally defined $\Sigma_0$ 
function symbols is a partial $\Pi$-algebra $A$ where 
(i) the reduct $A_{|\Pi_0}$ of $A$ to the signature $\Pi$ is a 
model of ${\cal T}_0$ (in the classical sense, i.e.\ all operations in 
$\Sigma_0$ are total); 
(ii) $A$ satisfies (in the Evans sense) all clauses in ${\cal K}$.  
}
A partial $\Pi$-algebra $A$ is a {\em weak partial model} of 
${\cal T}_1$ with totally defined $\Sigma_0$-function symbols if 
(i) $A_{|\Pi_0}$  is a model of ${\cal T}_0$ and 
(ii)
$A$ weakly satisfies all clauses in ${\cal K}$.  
\end{defi}

\noindent
If the base theory ${\cal T}_0$ and its signature are 
clear from the 
context, we will refer to 
{\em weak partial models} of ${\cal T}_1$. 
We will use the following notation: 

\begin{itemize}
\ignore{\item ${\sf PMod}(\Sigma_1, {\cal T}_1)$ the class of 
all partial models of 
${\cal T}_1$ in which the functions in $\Sigma_1$ are partial, and all other 
function symbols are total; 
} 
\item ${\sf PMod_w}({\Sigma_1}, {\cal T}_1)$ is  
the class of all weak 
partial models of ${\cal T}_1$ in which the 
$\Sigma_1$-functions are partial and all the other function symbols are total;
\item ${\sf PMod^f_w}({\Sigma_1}, {\cal T}_1)$ is the class of all finite weak 
partial models of ${\cal T}_1$ in which the 
$\Sigma_1$-functions are partial and all the other function symbols are total;
\item ${\sf PMod^{fd}_w}({\Sigma_1}, {\cal T}_1)$ is the class of all weak 
partial models of ${\cal T}_1$ in which the 
$\Sigma_1$-functions are partial and their definition domain is a finite set, 
and all the other function symbols are total;
\item ${\sf Mod}({\cal T}_1)$ denotes the class of all models of ${\cal T}_1$ 
in which all functions in $\Sigma_0 \cup \Sigma_1$ are totally defined.
\end{itemize}

\subsection{Embeddability}
For theory extensions 
${\cal T}_0 \subseteq  {\cal T}_1 = {\cal T}_0 \cup {\cal K}$,
where ${\cal K}$ is a set of clauses, 
we consider the following 
condition:

\medskip
\begin{tabular}{ll}
\ignore{${\sf (Emb)}$ $\quad$ & Every $A \in {\sf PMod}({\Sigma_1}, {\cal T}_1)$ 
weakly embeds into a total model of ${\cal T}_1$. \\}
${\sf (Emb_w)}$ $\quad$ & Every $A \in {\sf PMod_w}({\Sigma_1}, {\cal T}_1)$ 
weakly embeds into a total model of ${\cal T}_1$. 
\end{tabular}

\medskip
\noindent 
We also define a stronger notion of embeddability, which we call 
{\em completability}: 

\medskip
\noindent 
\begin{tabular}{ll}
$\quad {\sf (Comp_w)}$ \quad & Every  $A \in {\sf PMod_w}({\Sigma_1}, {\cal T}_1)$ 
weakly embeds into a total model $B$ of ${\cal T}_1$ \\
& such that $A_{|\Pi_0}$ and $B_{|\Pi_0}$ are isomorphic. \\
\end{tabular}

\medskip
\noindent 
Weaker conditions, which only refer to embeddability of {\em finite}
partial models,  will be denoted by 
${\sf (Emb^f_w)}$, resp.\ ${\sf (Comp^f_w)}$. Conditions which refer to 
embeddability of partial models in ${\sf PMod^{fd}_w}({\Sigma_1}, {\cal T}_1)$ 
will be denoted by 
${\sf (Emb^{fd}_w)}$, resp.\ ${\sf (Comp^{fd}_w)}$.

\section{Locality}
\label{sect:local-ext}

The notion of {\em local theory} was introduced by 
Givan and McAllester \cite{GivanMcAllester92,McAllester93}.
\begin{defi}[Local theory]
A local theory is a  set of Horn clauses 
${\cal K}$ such that, for any  ground Horn clause $C$, 
${\cal K} \models C$ only if 
already ${\cal K}[C] \models C$ (where ${\cal K}[C]$ is the set of 
instances of ${\cal K}$ in which all terms are subterms of 
ground terms in either ${\cal K}$ or ${\cal C}$).
\end{defi}
The notion of locality in {\em equational} theories was studied by 
Ganzinger \cite{Ganzinger-01-lics}, who also related it 
to a semantical property, namely embeddability of partial algebras into 
total algebras. In \cite{GSW04,GSW-iandc-06,Sofronie-cade-05} 
the notion of locality for Horn 
clauses is extended to the notion of {\em local extension} of a base theory. 

\medskip
\noindent Let ${\cal K}$ be a set of clauses in the signature 
$\Pi = (\Sigma_0 \cup \Sigma_1, {\sf Pred})$.
In what follows, when we refer to sets $G$ of ground clauses 
we assume that they are in the signature 
$\Pi^c = (\Sigma \cup \Sigma_c, {\sf Pred})$, 
where $\Sigma_c$ is a set of new constants. 
If $\Psi$ is a set of ground $\Sigma_0 \cup \Sigma_1 \cup \Sigma_c$-terms, 
we denote by ${\cal K}_{\Psi}$ the set of all instances of ${\cal K}$ 
in which all terms starting with a $\Sigma_1$-function symbol 
are ground terms in the set $\Psi$.
If $G$ is a set of ground clauses and 
$\Psi = {\sf st}({\cal K}, G)$ is the set of ground subterms occurring 
in  either ${\cal K}$ or $G$ then we write 
${\cal K}[G] := {\cal K}_{\Psi}$.

\medskip
\noindent We will focus on the following type of locality of 
a theory extension 
${\cal T}_0 \subseteq {\cal T}_1$, where 
${\cal T}_1 = {\cal T}_0 \cup {\cal K}$ with ${\cal K}$ a set of (universally
quantified) clauses: 

\medskip
\noindent
\begin{tabular}{ll}
${\sf (Loc)}$ $~~$ & For every set $G$ of ground clauses ${\cal T}_1 \cup G \models \perp$ iff ${\cal T}_0 \cup {\cal K}[G] \cup G$ has \\
& no weak partial model in which all terms in ${\sf st}({\cal K}, G)$ are defined. \\[1ex]
\end{tabular}

\medskip
\noindent 
A weaker notion ${\sf (Loc^f)}$
can be defined if we require that the 
respective conditions hold only for {\em finite} sets $G$ of ground clauses.
An intermediate notion of locality ${\sf (Loc^{fd})}$
can be defined if we require that the 
respective conditions hold only for sets $G$ of ground clauses containing 
only a finite set of terms starting with a function 
symbol in $\Sigma_1$. 

\begin{defi}[Local theory extension]
An extension ${\cal T}_0 \subseteq {\cal T}_1$ is 
{\em local} 
if it satisfies condition ${\sf (Loc^f)}$.
\end{defi}
A local theory \cite{Ganzinger-01-lics} 
is a local extension of the empty theory.

\section{Locality and embeddability}
\label{embed}

There is a strong link between locality of a theory extension and 
embeddability of partial models into total ones. Links between 
{\em locality of a theory} and {\em embeddability} were established 
by Ganzinger in \cite{Ganzinger-01-lics}. We show that 
similar results can be obtained also for {\em local theory extensions}.

\medskip 
\noindent 
In what follows we say that a non-ground clause is $\Sigma_1$-{\em flat} 
if function symbols (including constants) do not occur 
as arguments of function symbols in $\Sigma_1$.
A $\Sigma_1$-flat non-ground clause is called $\Sigma_1$-{\em linear} 
if whenever a 
variable occurs in two terms in the clause 
which start with function symbols in $\Sigma_1$, 
the two terms
are identical, and if 
no term which starts with a function 
in $\Sigma_1$ contains two occurrences of the same variable.

\subsection{Locality implies embeddability}

\noindent 
We first show that for sets of $\Sigma_1$-flat clauses locality 
implies embeddability. This generalizes results presented in the 
case of local theories in \cite{Ganzinger-01-lics}. 
\begin{thm}
Assume that ${\cal K}$ is a family of $\Sigma_1$-flat clauses 
in the signature $\Pi$. 
\begin{itemize}
\item[(1)] If the extension ${\cal T}_0 \subseteq {\cal T}_1 := {\cal T}_0 \cup {\cal K}$
satisfies ${\sf (Loc)}$ then it satisfies ${\sf (Emb_w)}$. 
\item[(2)] If the extension ${\cal T}_0 \subseteq {\cal T}_1 := {\cal T}_0 \cup {\cal K}$
satisfies ${\sf (Loc^f)}$ then it satisfies ${\sf (Emb^f_w)}$. 
\item[(3)] If the extension ${\cal T}_0 \subseteq {\cal T}_1 := {\cal T}_0 \cup {\cal K}$
satisfies ${\sf (Loc^{fd})}$ then it satisfies ${\sf (Emb^{fd}_w)}$. 
\item[(4)] If ${\cal T}_0$ is compact and 
the extension ${\cal T}_0 \subseteq {\cal T}_1$
satisfies ${\sf (Loc^f)}$,  then ${\cal T}_0 \subseteq {\cal T}_1$ satisfies 
${\sf (Emb_w)}$.   
\end{itemize}
\label{locality-implies-embedding}
\end{thm}
\Proof We prove (4) and show how the proof can be 
changed to provide proofs for (1), (2) and (3).
Let $A$ be a partial  $\Pi$-algebra with 
totally defined $\Sigma_0$-functions, which is a model of ${\cal T}_0$
and weakly satisfies ${\cal K}$. 
Let 
\begin{eqnarray*}
\Delta(A) & = & \{ f(a_1, \dots a_n) \approx a \mid \text{ if } 
f_A(a_1, \dots, a_n) \text{ is defined and equal to } a \} \\
          & & \cup \{ f(a_1, \dots a_n) \not\approx a \mid \text{ if } 
f_A(a_1, \dots, a_n) \text{ is defined and not equal to } a \} \\
          & & \cup \{ P(a_1, \dots, a_n) \mid P \in {\sf Pred} \text{ and } (a_1, \dots, a_n) \in P_A \} \\
          & & \cup \{ \neg P(a_1, \dots, a_n) \mid P \in {\sf Pred} 
\text{ and } (a_1, \dots, a_n) \not\in P_A \} \cup \bigwedge_{a \neq a', a, a' \in A} a \not\approx a' 
\end{eqnarray*}

\noindent 
We prove that ${\cal T}_0 \cup {\cal K} \cup \Delta(A)$ is consistent, 
where the elements of $A$
are regarded as new constants.
Assume ${\cal T}_0 \cup {\cal K} \cup \Delta(A) \models \perp$. 
By compactness of ${\cal T}_0$, 
${\cal T}_0 \cup {\cal K} \cup \Gamma \models \perp$, for some 
finite subset $\Gamma$ of $\Delta(A)$. 
%
%
We know that $A$ is a model of ${\cal T}_0$.
Every term starting with a function symbol in $\Sigma_1$ 
contained in the clauses in ${\cal K}[\Gamma]$ is 
either a ground (subterm of a) term occurring in $\Gamma$ 
(and, hence, a constant $a \in A$, or a 
term $f(a_1, \dots, a_n)$, where $f_A(a_1, \dots, a_n)$ is defined),
or is a ground subterm in ${\cal K}$, i.e.\ a constant, and hence, 
again defined in $A$.  
Therefore, all terms occurring in the clauses in 
${\cal K}[\Gamma]$ are defined in $A$, 
so $A$ satisfies all these clauses, i.e.\ $A$ is a model of 
${\cal T}_0 \cup {\cal K}[\Gamma]$. 
Since $\Delta(A)$ is obviously true in $A$ and $\Gamma \subseteq \Delta(A)$, 
$A$ is a partial model of 
${\cal T}_0 \cup {\cal K}[\Gamma] \cup \Gamma$, in which all ground terms
occurring in ${\cal K}$ or $\Gamma$ are defined. 
This contradicts the fact that ${\cal T}_1$ is a local extension of 
${\cal T}_0$.
Hence, the assumption that ${\cal T}_0 \cup {\cal K} \cup \Delta(A) 
\models \perp$
was false, so ${\cal T}_0 \cup {\cal K} \cup \Delta(A)$ has a model $A'$
in which, therefore, $A$ weakly embeds. 

\smallskip
\noindent 
(1) If ${\sf (Loc)}$ holds then 
we can choose $\Gamma = \Delta(A)$. 
(2) If $A$ is finite 
we can choose $\Gamma = \Delta(A)$, so 
the compactness of ${\cal T}_0$ is not needed.
(3) If all functions in $\Sigma_1$ have 
a finite domain of definition in $A$, then $\Delta(A)$ contains only finitely
many terms starting with a $\Sigma_1$-function. Therefore also in 
this case we can choose $\Gamma = \Delta(A)$.  \QED

\subsection{Embeddability implies locality}

\noindent 
Conversely, embeddability implies locality. 
The following results appear in \cite{Sofronie-cade-05} 
and \cite{sofronie-ihlemann-ismvl-07}. 
This result allows to give several examples of local theory extensions.

\begin{thm}[\cite{Sofronie-cade-05,sofronie-ihlemann-ismvl-07}]
Let ${\cal K}$ be a set of $\Sigma$-flat and $\Sigma$-linear clauses.
\begin{itemize}
\item[(1)] 
If the extension ${\cal T}_0 \subseteq {\cal T}_1$ satisfies ${\sf (Emb_w)}$ then it satisfies ${\sf (Loc)}$.
\item[(2)] Assume that ${\cal T}_0$ is a locally 
finite universal theory, and that ${\cal K}$ contains only finitely 
many ground subterms. 
If the extension ${\cal T}_0 \subseteq {\cal T}_1$ satisfies 
${\sf (Emb^f_w)}$, then ${\cal T}_0 \subseteq {\cal T}_1$ satisfies 
${\sf (Loc^f)}$.  
\item[(3)] ${\cal T}_0 \subseteq {\cal T}_1$ satisfies 
${\sf (Emb^{fd}_w)}$. 
Then  ${\cal T}_0 \subseteq {\cal T}_1$ satisfies ${\sf (Loc^{fd})}$.
\end{itemize}
\label{rel-loc-embedding}
\end{thm}

\section{Examples of local theory extensions}
\label{examples}

We present several examples of theory extensions for which embedding 
conditions among those mentioned above hold and are thus local. 
For details cf.\ 
\cite{Sofronie-cade-05,Sofronie-ijcar-06,sofronie-ihlemann-ismvl-07}. 

\begin{description}
\item[Extensions with free functions.] 
Any extension ${\cal T}_0 \cup {\sf Free}(\Sigma)$ 
of a theory ${\cal T}_0$ with a set $\Sigma$ of free function 
symbols satisfies condition ${\sf (Comp_w)}$.

\medskip
\item[Extensions with selector functions.] 
Let ${\cal T}_0$ be a theory with signature 
$\Pi_0 = (\Sigma_0, {\sf Pred})$, let $c \in \Sigma_0$ 
with arity $n$, and let 
$\Sigma_1 = \{ s_1, \dots, s_n \}$ consist of $n$ unary 
function symbols.
Let ${\cal T}_1 = {\cal T}_0 \cup {\sf Sel}_c$ 
(a theory with signature $\Pi = (\Sigma_0 \cup \Sigma_1, {\sf Pred})$) be the 
extension of ${\cal T}_0$ with the set 
${\sf Sel}_c$ of clauses below.
Assume that ${\cal T}_0$ satisfies the (universally quantified) formula 
${\sf Inj_c}$ (i.e.\ $c$ is injective in ${\cal T}_0$) then the 
extension ${\cal T}_0 \subseteq {\cal T}_1$ 
satisfies condition ${\sf (Comp_w)}$ \cite{Sofronie-cade-05}. 
{\small 
\begin{eqnarray*}
({\sf Sel}_c) & \quad \quad & s_1(c(x_1, \dots, x_n))  \approx  x_1  \\[-0.5ex]
& & \cdots \\[-0.5ex]
& & s_n(c(x_1, \dots, x_n)) \approx x_n \\
& & x \approx c(x_1, \dots, x_n)  \rightarrow  c(s_1(x), \dots, s_n(x)) \approx x \\
({\sf Inj_c}) & \quad \quad & c(x_1, \dots, x_n) \approx c(y_1, \dots, y_n) \rightarrow (\bigwedge_{i = 1}^n x_i \approx y _i)
\end{eqnarray*}
}

\medskip
\item[Extensions with functions satisfying general monotonicity conditions.]
In \cite{Sofronie-cade-05} and 
\cite{sofronie-ihlemann-ismvl-07} we analyzed extensions with monotonicity 
conditions for an $n$-ary function $f$ 
w.r.t.\ a subset $I \subseteq \{ 1, \dots, n \}$ of its arguments: 
$${ ({\sf Mon}^I_f) ~  {\bigwedge_{i \in I}} x_i {\leq_i} y_i {\wedge} {\bigwedge_{i \not\in I}} x_i {=} y_i {\rightarrow} f(x_1,.., x_n) {\leq} f(y_1,.., y_n).}$$
\noindent 
Here, ${\sf Mon}^{\emptyset}_f$ is equivalent to the congruence axiom for 
$f$. If $I = \{ 1, \dots, n \}$ we speak of monotonicity in all arguments;
we denote ${\sf Mon}^{\{1,\dots,n\}}_f$ by ${\sf Mon}_f$. 
Monotonicity in some arguments and antitonicity in other arguments is modeled
by considering functions $f : \prod_{i \in I} P_i^{\sigma_i} \times 
\prod_{j \not\in I} P_j \rightarrow P$ with 
$\sigma_i \in \{ -, + \}$, 
where $P_i^+ = P_i$ and $P_i^- = P_i^{\partial}$, the order dual of the poset 
$P_i$.
The corresponding axioms are denoted by 
${\sf Mon}^{\sigma}_f$, where for 
$i \in I$, $\sigma(i) = \sigma_i  \in \{ -, + \}$, and for 
$i \not\in I$, $\sigma(i) = 0$.
The following hold \cite{Sofronie-cade-05,sofronie-ihlemann-ismvl-07}: 

\begin{enumerate}
\item Let ${\cal T}_0$ be a class of (many-sorted) 
bounded semilattice-ordered $\Sigma_0$-structures. Let $\Sigma_1$ be disjoint 
from $\Sigma_0$ 
and  ${\cal T}_1 = {\cal T}_0 {\cup} \{ {\sf Mon}^{\sigma}(f) {\mid} 
f \in \Sigma_1 \}$.  Then the extension ${\cal T}_0 \subseteq {\cal T}_1$ 
satisfies  $({\sf Comp^{fd}_w})$, hence is local. 
\item Any extension of the theory of posets with 
functions in a set $\Sigma_1$
satisfying $\{ {\sf Mon}^{\sigma}_f \mid f \in \Sigma_1 \}$ 
satisfies condition $({\sf Emb_w}),$ hence is local.
\end{enumerate}

This provides us with a large number of concrete examples. For instance   
the extensions 
with functions satisfying monotonicity axioms ${\sf Mon}^{\sigma}_f$ 
of the following (possibly many-sorted) classes of algebras are local:

\begin{itemize}
\item any class of algebras with a bounded (semi)lattice reduct, a 
bounded distributive lattice reduct, or a Boolean algebra reduct 
($({\sf Comp^{fd}_w})$ holds); 
\item any extension of a class of algebras with a semilattice reduct, 
a (distributive) lattice reduct, or a
Boolean algebra reduct, with monotone functions 
into an infinite numeric domain
($({\sf Comp^{fd}_w})$ holds);
\item $\cal T$, the class of totally-ordered sets; 
$\cal{DO}$, the theory of dense totally-ordered sets  
($({\sf Comp^{fd}_w})$ holds);  
\item the class ${\cal P}$ of partially-ordered sets 
($({\sf Emb_w})$ holds). 
\end{itemize}

Similarly, it can be proved that any extension of 
the theory of reals (integers) 
with functions satisfying $Mon^{\sigma}_f$ into a fixed infinite numerical 
domain is local (condition $({\sf Comp^{fd}_w})$ holds). 

\medskip
\item[Boundedness conditions.] 
Any extension of a theory for which $\leq$ is reflexive 
with functions satisfying $({\sf Mon}^{\sigma}_f)$ and boundedness 
$({\sf Bound}^t_{f})$  conditions 
is local \cite{Sofronie-ijcar-06,sofronie-ihlemann-ismvl-07}. 

\smallskip
$({\sf Bound}^t_{f}) \quad  \quad \forall x_1, \dots, x_n (f(x_1, \dots, x_n) \leq t(x_1, \dots, x_n))$ 

\smallskip 
\noindent 
where $t(x_1, \dots, x_n)$ is a term in the base signature $\Pi_0$
with variables among $x_1, \dots, x_n$ (such that in any model the 
associated function has the same monotonicity as $f$).

\medskip
Similar results can be given for {\em guarded monotonicity conditions}
with mutually disjoint guards \cite{Sofronie-ijcar-06}. 
 
\medskip
\item[Extensions with Lipschitz functions.]
The extension ${\mathbb R} \subseteq {\mathbb R} \cup ({\sf L_f^{\lambda}})$ 
of ${\mathbb R}$ with a unary function which is 
$\lambda$-Lipschitz in a point $x_0$ (for $\lambda > 0$)
satisfies condition ${\sf (Comp_w)}$.

$$({\sf L}_f^{\lambda}) \quad \quad 
\forall x ~ | f(x) - f(x_0) | \leq \lambda \cdot | x - x_0|$$
\end{description}
The results described before can easily be extended 
to a many-sorted framework. Therefore various additional examples
of (many-sorted) theory extensions related to data structures
can be given cf.\ e.g.\ \cite{Sofronie-verify-06}. 

\section{Combinations of local extensions satisfying $({\sf Comp_w})$} 
\label{comp}

In this and the following sections we study the locality of combinations of 
local theory extensions. In the light of the results in Section~\ref{embed}
we concentrate on studying which embeddability properties are preserved 
under combinations of theories. For the sake of simplicity, in what 
follows we consider only conditions $({\sf Emb_w})$ and $({\sf Comp_w})$. 
Analogous results can be given for conditions 
$({\sf Emb^{f}_w})$, $({\sf Comp^{f}_w})$, resp.\ 
$({\sf Emb^{fd}_w})$, $({\sf Comp^{fd}_w})$ and combinations thereof. 

\bigskip
\noindent 
We start with a simple case of combinations of local extensions of a base 
theory: we consider the situation when both components 
satisfy the embeddability condition $({\sf Comp_w})$. 
We first analyze the simple case of combinations of local extensions of 
a base theory ${\cal T}_0$ by means of sets of mutually disjoint function 
symbols. Then some results on combining extensions with non-disjoint 
sets of function symbols are discussed.

\begin{thm}
Let ${\cal T}_0$ be a first-order theory
with signature $\Pi_0 = (\Sigma_0, {\sf Pred})$ 
and 
${\cal T}_1 = {\cal T}_0 \cup {\cal K}_1$ and 
${\cal T}_2 = {\cal T}_0 \cup {\cal K}_2$
two extensions of ${\cal T}_0$ with signatures 
$\Pi_1 = (\Sigma_0 \cup \Sigma_1, {\sf Pred})$ and 
$\Pi_2 = (\Sigma_0 \cup \Sigma_2, {\sf Pred})$, respectively. 
Assume that both extensions ${\cal T}_0 \subseteq {\cal T}_1$ 
and ${\cal T}_0 \subseteq {\cal T}_1$ satisfy condition $({\sf Comp_w})$, 
and that $\Sigma_1 \cap \Sigma_2 = \emptyset$. 
Then the extension 
${\cal T}_0 \subseteq {\cal T} = {\cal T}_0 \cup {\cal K}_1 \cup {\cal K}_2$
satisfies condition $({\sf Comp_w})$. If, additionally, 
in ${\cal K}_i$ all terms 
starting with a function symbol in $\Sigma_i$ are flat and linear, 
for $i = 1, 2$,  
then the extension is local. 
\label{combinations-satisfying-comp}
\end{thm}
\Proof 
Let $P \in {\sf PMod_w}(\Sigma_1 \cup \Sigma_2, {\cal T})$. 
Then $P_{| \Pi_1} \in {\sf PMod_w}(\Sigma_1, {\cal T}_1)$, hence 
$P_{| \Pi_1}$ weakly embeds into a total model $B$ of ${\cal T}_1$, 
such that  $P_{|\Pi_0}$ and $B_{|\Pi_0}$ are isomorphic. 
Let $i : P_{|\Pi_0} \rightarrow B_{|\Pi_0}$ be the isomorphism between these
two $\Pi_0$-structures. 
We use the isomorphism $i$ to transfer also the 
$\Sigma_2$-structure from  $P$ to $B$. 
That is, for every $f \in \Sigma_2$ with arity $n$, 
and every 
$b_1, \dots, b_n \in B$, we define:
$$
f_B(b_1, \dots, b_n) = \left \{ \begin{array}{ll}
i(f_P(i^{-1}(b_1), \dots, i^{-1}(b_n)))~~~~ & 
\text{ if } f_P(i^{-1}(b_1), \dots, i^{-1}(b_n))  \\ 
& \text{ is defined in } P \\[1ex]
\text{undefined} & \text{ otherwise} \end{array} \right.
$$
With these definitions of $\Sigma_2$-functions, 
$B_{| \Pi_2} \in {\sf PMod_w}(\Sigma_2, {\cal T}_2)$. 
Therefore, $B_{| \Pi_2}$ weakly embeds into a total model $C$ 
of ${\cal T}_1$, 
such that  $B_{|\Pi_0}$ and $C_{|\Pi_0}$ are isomorphic. 
Let $j : B_{|\Pi_0} \rightarrow C_{|\Pi_0}$ be the isomorphism between these
two structures. We use this isomorphism to transfer, as explained above, 
the (total) $\Sigma_1$-structure from  $B$ to $C$.
The algebra $A$ obtained this way from $C$ is a total model of ${\cal T}$, and 
$j \circ i : P_{|\Pi_0} \rightarrow A_{|\Pi_0}$ is an isomorphism. 
Thus, the extension 
${\cal T}_0 \subseteq {\cal T} = {\cal T}_0 \cup {\cal K}_1 \cup {\cal K}_2$
satisfies condition $({\sf Comp_w})$. 
The last claim is an immediate consequence of Theorem~\ref{rel-loc-embedding}. 
\QED

\begin{ex}
{\em The following combinations of theories (seen as extensions of 
a first-order theory ${\cal T}_0$) satisfy condition $({\sf Comp_w})$
(or in case (4) condition $({\sf Comp^{fd}_w})$): 

\begin{enumerate}
\item[(1)] ${\cal T}_0 \cup {\sf Free}(\Sigma_1)$ and
${\cal T}_0 \cup {\sf Sel}_c$ if ${\cal T}_0$ is a theory and $c \in \Sigma_0$ is injective in ${\cal T}_0$. 
\item[(2)] ${\mathbb R} \cup {\sf Free}(\Sigma_1)$ and ${\mathbb R} \cup {\sf Lip}^{\lambda}_c(f)$, where $f \not\in \Sigma_1$. 
\item[(3)] ${\mathbb R} \cup {\sf Lip}^{\lambda_1}_{c_1}(f)$  and ${\mathbb R} \cup {\sf Lip}^{\lambda_2}_{c_2}(g)$, where $f \neq g$. 
\item[(4)] ${\cal T}_0 \cup {\sf Free}(\Sigma_1)$ and 
${\cal T}_0 \cup {\sf Mon}^{\sigma}_f$, where 
$f \not\in \Sigma_1$ has arity $n$, 
$\sigma : \{ 1, \dots, n \} \rightarrow \{ -1, 1, 0 \}$, if 
${\cal T}_0$ is, e.g.,   
a theory of algebras with a bounded semilattice reduct.
\end{enumerate}
}
\end{ex}

\noindent 
A more general result holds, which allows to prove locality 
also for extensions which share non-base function symbols.
\begin{thm}
Let ${\cal T}_0$ be an arbitrary first-order theory, and 
${\cal T}_1 = {\cal T}_0 \cup {\cal K}_1$ and 
${\cal T}_2 = {\cal T}_0 \cup {\cal K}_2$
two extensions of ${\cal T}_0$ with functions in $\Sigma_1$ and $\Sigma_2$ 
respectively, which satisfy condition 
$({\sf Comp_w})$. 
Assume that there exists a set ${\cal K}$ of clauses in signature 
$\Sigma_0 \cup \Sigma$, where $\Sigma = \Sigma_1 \cap \Sigma_2 \subset \Sigma_i$, $i = 1,2$, 
such that every model of ${\cal T}_0 \cup {\cal K}_i$ is a model of 
${\cal T}_0 \cup {\cal K}$ for $i = 1, 2$. 
Then the extension  
${\cal T}_0 \cup {\cal K} \subseteq 
({\cal T}_0 \cup {\cal K}) \cup {\cal K}_1 \cup {\cal K}_2$
again satisfies condition $({\sf Comp_w})$ and hence is a local extension.
\label{combinations-satisfying-comp-non-disj}
\end{thm}
\Proof Note that if 
${\cal T}_0 \subseteq {\cal T}_0 \cup {\cal K}_i$ satisfies condition 
$({\sf Comp_w})$ then the extension ${\cal T}_0 \cup {\cal K} \subseteq 
 ({\cal T}_0 \cup {\cal K}) \cup {\cal K}_i$ also satisfies condition 
$({\sf Comp_w})$. The conclusion now follows from 
Theorem~\ref{combinations-satisfying-comp}, taking into account the fact that 
the signatures $(\Sigma_1 \backslash \Sigma)$ and 
$(\Sigma_2 \backslash \Sigma)$ are disjoint. 
\QED

\begin{ex}
{\em The following theory extensions satisfy condition $({\sf Comp_w})$: 

\begin{enumerate}
\item[(1)] ${\cal T}_0 \cup {\sf Free}(\Sigma) \subseteq 
({\cal T}_0 \cup {\sf Free}(\Sigma \cup \Sigma_1)) \cup 
({\cal T}_0 \cup {\sf Free}(\Sigma) \cup {\sf Sel}_c)$, provided that   
${\cal T}_0$ is a theory containing an injective  function $c$.
\item[(2)] ${\mathbb R} \cup {\sf Free}(f) \subseteq 
({\mathbb R} \cup {\sf Mon}_f \cup {\sf Mon}_g) \cup 
({\mathbb R} \cup {\sf Free}(f) \cup {\sf Lip}^{\lambda}_c(h))$, 
where $f, g, h$ are different function symbols. 
\item[(3)] ${\mathbb R} \cup {\sf Lip}^{\lambda_2}_{c}(f) \subseteq ({\mathbb R} \cup {\sf Lip}^{\lambda_1}_{c}(f) \cup {\sf Mon}(g)) \cup 
({\mathbb R} \cup {\sf Lip}^{\lambda_2}_{c}(f) \cup {\sf Free}(h))$, 
where $f, g, h$ are different function symbols and $\lambda_1 \leq \lambda_2$.
\end{enumerate}
}
\end{ex}
\Proof Immediate consequences of 
Theorem~\ref{combinations-satisfying-comp-non-disj}. (1) is obvious; 
for (2) note that every model of ${\mathbb R} \cup {\sf Mon}_f \cup 
{\sf Mon}_g$
is a model of ${\mathbb R} \cup {\sf Free}(f)$; for (3) note that, as
$\lambda_1 \leq \lambda_2$,  
every model of 
${\mathbb R} \cup {\sf Lip}^{\lambda_1}_{c}(f) \cup {\sf Mon}(g)$ is a model 
of ${\mathbb R} \cup {\sf Lip}^{\lambda_2}_{c}(f)$. \QED

\section{More general combinations of local theory extensions}
\label{emb}

The result above can be extended to the more general situation in which 
one of the extensions, say ${\cal T}_0 \subseteq {\cal T}_1 = {\cal T}_0 \cup {\cal K}_1$,  satisfies condition $({\sf Emb_w})$ and the other extension 
${\cal T}_0 \subseteq {\cal T}_2 = {\cal T}_0 \cup {\cal K}_2$
satisfies condition $({\sf Comp_w})$, or if both extensions satisfy condition 
$({\sf Emb_w})$. The natural analogon of the proof of 
Theorem~\ref{combinations-satisfying-comp} would be the following: Start with a partial model $P$ of ${\cal T}_0 \cup 
{\cal K}_1 \cup {\cal K}_2$; extend it, using property 
$({\sf Emb_w})$, to a total model $A$ of ${\cal T}_1$.  
The technical problem which occurs when we now try to use the embedding 
property for ${\cal T}_2$ is that we need to 
be sure that $A$ remains also a partial model of ${\cal T}_2$, 
with the operations inherited from $P$. 
Unfortunately this may not always be the case, as shown below.  
\begin{ex}
{\em Let $\Pi_0 = (\{ f \}, {\sf Pred})$ and let 
${\cal T}_0$ be a $\Pi_0$-theory. 
Let ${\cal T}_1 = {\cal T}_0 \cup {\cal K}_1$, and ${\cal T}_2 = {\cal T}_0 \cup {\cal K}_2$ be two theories over extensions of $\Pi_0$ 
with function symbols in $\Sigma_1, \Sigma_2$. 
Assume that 
$\Sigma_2 = \{ g \}$, $\Sigma_1 \cap \Sigma_2 = \emptyset$, and 
${\cal K}_2 = \{ x  = f(x) \rightarrow g(y) = y \}$ 
($f$ and $g$ are unary function symbols).

Let $P = (\{ a, b \}, f_P, g_P, \{ \sigma_P \}_{\sigma \in \Sigma_1})$ 
be a partial 
algebra, where $f_P$ is total with $f_P(a) = b$ and $f_P(b) = a$; 
$g_P(a) = b$ and $g_P(b)$ is undefined.
$P$ weakly satisfies ${\cal K}_2$ because the premise 
of the clause in ${\cal K}_2$ is always false in $P$.
Assume that $P$ weakly embeds into a total model $A$ of ${\cal T}_1$
via a $\Pi_1$-embedding $h : P \hookrightarrow A$, and that $A$ 
contains an element $c \not\in \{ h(a), h(b) \}$, such that 
$f_A(c) = c$. $A$ ``inherits'' the $\Sigma_2$-operation $g$ from $P$ via $h$, 
in the sense that  
we can define $g_A(h(a)) := h(g_P(a)) = h(b)$ and assume that  
$g_A$ is undefined in rest.
However, with the $\Sigma_2$-operation defined this way 
$A$ does not weakly satisfy ${\cal K}_2$. 
Let $\beta : X \rightarrow A$ with 
$\beta(x) = c$ and $\beta(y) = h(a)$. 
$(A, \beta)$ does not weakly satisfy the clause in 
${\cal K}_2$, since: 
\begin{eqnarray*}
\beta(f(x)) & = & f_A(\beta(x)) = f_A(c) = c, \text{ whereas } \\
\beta(g(y)) & = & g_A(\beta(y)) = g_A(h(a)) = h(g_P(a)) = h(b) \neq h(a) = \beta(y).
\end{eqnarray*} 
This happens because the variable $x$ in the clause in ${\cal K}_2$ 
does not occur below any function symbol in $\Sigma_2$. 
}
\end{ex}
In what follows we identify conditions which ensure 
that an extension $A$ of a partial algebra $P$ which weakly satisfies 
${\cal K}_2$ remains a partial model of ${\cal K}_2$ 
with the $\Sigma_2$-operations inherited from $P$, 

\subsection{Preservation of truth under extensions}

\begin{lem}
Let ${\cal T}_0$ be a theory with
signature $\Pi_0 = (\Sigma_0, {\sf Pred})$, and 
let ${\cal T}_0 \subseteq {\cal T} := {\cal T}_0 \cup {\cal K}$ 
be a theory extension by means of a set ${\cal K}$ of $\Sigma$-flat 
clauses over the signature $\Pi = (\Sigma_0 \cup \Sigma, {\sf Pred})$. 
Assume that for each clause $C$ of ${\cal K}$ all variables in 
$C$ occur below some $\Sigma$-function symbol. 

Let $P \in {\sf PMod_w}(\Sigma, {\cal T})$, $A \in {\sf Mod}({\cal T}_0)$, 
and $h : P \hookrightarrow  A$ be a $\Pi_0$-embedding. 
Then a partial $\Sigma$-structure can be defined on $A$ such that 
$A$ weakly satisfies ${\cal K}$, and $h$ is a weak 
$\Pi$-embedding. 
\label{lemma:ext-struct}
\end{lem}
\Proof For every $a_1, \dots, a_n \in A$ and every $f \in \Sigma$
define 
$$f_A(a_1, \dots, a_n) := \left\{ \begin{array}{ll}
a & \text{ if } \exists p_1, \dots, p_n \in P 
\text{ such that all } a_i = h(p_i), \\
& f_P(p_1, \dots, p_n) \text{ is defined in } P, \\
& \text{and } a = h(f_P(p_1, \dots, p_n)) \\ 
\text{ undefined } & \text{ otherwise.}
\end{array} \right. $$ 
As $h$ is injective, $f_A$ is well-defined. By hypothesis, 
$h$ is a $\Pi_0$-embedding. 
With the definition of operations in $\Sigma$ given above, 
$h$ is also a weak $\Sigma$-homomorphism. 
Let $p_1, \dots, p_n \in P$ and $f \in \Sigma$ be such that 
$f_P(p_1, \dots, p_n)$ is defined. Then, by the definition of $f_A$,
$f_A(h(p_1), \dots, h(p_n))$ is defined and equal to 
$h(f_P(p_1, \dots, p_n))$. 
 
We now prove that with the operations defined as shown before 
$A$ weakly satisfies ${\cal K}$. Let $C \in {\cal K}$ and let 
$\beta : X \rightarrow A$ be an 
assignment of elements in $A$ to the variables in $C$.
 Assume that for every term $t$ occurring in $C$, $\beta(t)$ is defined 
in $A$ (otherwise, due to the definition of weak satisfiability, 
$(A, \beta) \models_w C$ trivially). 
 In order to show that $(A, \beta) \models_w C$, 
 we construct 
 an assignment $\alpha$ of elements in $P$ to the variables in $C$, 
 and use the fact that $(P, \alpha) \models_w C$.

 Let $t = f(t_1, \dots, t_k)$ be an arbitrary term occurring in $C$,   
with $f \in \Sigma$. 
 As $\beta(t)$ is defined, $f_A(\beta(t_1), \dots, \beta(t_k))$ is defined 
 in $A$, hence there exist $p_{1}, \dots, p_{k} \in P$ such that 
 $h(p_{i}) = \beta(t_i)$, $f_P(p_{1}, \dots, p_{k})$ is defined, and 
 $f_A(\beta(t_1), \dots, \beta(t_k)) = h(f_P(p_{1}, \dots, p_{n}))$. 
As all clauses in ${\cal K}$ are ${\Sigma}$-flat, all terms $t_i$ 
are variables.
 In this way we can associate with every variable $x$ occurring as argument in 
a term $f(t_1, \dots, t_n)$ of $C$ with $f \in \Sigma$
 an element $p_x \in P$ such that $h(p_x) = \beta(x)$. 
 Assume that for some such (variable) subterm $x$, two elements of $P$, 
 say $p_x$ and $q_x$, can be associated in this way. Then 
 $h(p_x) = \beta(x) = h(q_x)$, and the injectivity of $h$ guarantees that 
 $p_x = q_x$. 
This shows that an assignment 
$\alpha : X \rightarrow P$ can be defined, such that 
for all variables in $C$ 
occurring below a function symbol in $\Sigma$ 
(hence for all variables in $C$)  $\alpha(x) := p_x$. 
It is easy to see that for every term $t$ occurring in $C$, 
$h(\alpha(t)) = \beta(t)$.
As 
$(P, \alpha) \models C$ and $h$ is a weak $\Pi$-embedding 
it follows that $(A, \beta) \models C$. \QED

\medskip
\noindent The result above will be applied in 
Theorems~\ref{combinations-emb-comp} 
and~\ref{combine-disjoint} 
in the following form:

\begin{cor}
Let ${\cal T}_0$ be a first-order theory with 
signature $\Pi_0 = (\Sigma_0, {\sf Pred})$.
Let $\Sigma_1, \Sigma_2$ be two disjoint sets of function symbols, 
and let $\Pi_i = (\Sigma_0 \cup \Sigma_i, {\sf Pred})$, $i = 1, 2$, 
and $\Pi = (\Sigma_0 \cup \Sigma_1 \cup \Sigma_2, {\sf Pred})$. 
Let ${\cal K}_2$ be a set of $\Sigma_2$-flat clauses over 
$\Pi_2$. Assume that for each clause $C$ of ${\cal K}_2$ all variables in 
$C$ occur below some function symbol in $\Sigma_2$. 

Let $P$ be a partial 
$\Pi$-structure such that $P_{|\Pi_0}$ is a total 
model of ${\cal T}_0$, and  
$P$ weakly satisfies ${\cal K}_2$. 
Let $A$ be a total $\Pi_1$-structure, and  
let $h : P \hookrightarrow A$ be a weak 
$\Pi_1$-embedding. 
Then a partial $\Sigma_2$-structure can be defined on $A$ such that 
$A$ weakly satisfies ${\cal K}_2$, and $h$ is a weak 
$\Pi$-embedding. 
\label{cor:ext-struct}
\end{cor}

\subsection{Combining local extensions, one of which satisfies $({\sf Comp_w})$} 
We now analyze the case of combinations of theories in which 
one component satisfies condition $({\sf Comp_w})$ and the other 
component satisfies condition $({\sf Emb}_w)$. 
\begin{thm}
Let ${\cal T}_0$ be a first-order theory with
signature $\Pi_0 = (\Sigma_0, {\sf Pred})$, 
and let 
${\cal T}_1 = {\cal T}_0 \cup {\cal K}_1$ and 
${\cal T}_2 = {\cal T}_0 \cup {\cal K}_2$
be two extensions of ${\cal T}_0$ with signatures 
$\Pi_1 = (\Sigma_0 \cup \Sigma_1, {\sf Pred})$ and 
$\Pi_2 = (\Sigma_0 \cup \Sigma_2, {\sf Pred})$, respectively. 
Assume that:  
\begin{itemize}
\item[(1)] ${\cal T}_0 \subseteq {\cal T}_1$ satisfies condition 
$({\sf Comp_w})$, 
\item[(2)] ${\cal T}_0 \subseteq {\cal T}_2$ 
satisfies condition $({\sf Emb_w})$, 
\item[(3)] ${\cal K}_1$ is a set of $\Sigma_1$-flat clauses  in which 
all variables occur below a $\Sigma_1$-function. 
\end{itemize} 
Then the extension 
${\cal T}_0 \subseteq {\cal T}_0 \cup {\cal K}_1 \cup {\cal K}_2$
satisfies $({\sf Emb_w})$. If, additionally, 
in ${\cal K}_i$ all terms 
starting with a function symbol in $\Sigma_i$ are flat and linear, 
for $i = 1, 2$,  
then the extension is local. 
\label{combinations-emb-comp}
\end{thm}
\Proof 
Let $P \in {\sf PMod_w}(\Sigma_1 \cup \Sigma_2, {\cal T}_0 \cup {\cal K}_1 \cup {\cal K}_2)$. 
Then $P_{| \Pi_2} \in {\sf PMod_w}(\Sigma_2, {\cal T}_2)$, hence 
$P_{| \Pi_2}$ weakly embeds into a total model $B$ of ${\cal T}_2$.
By (3), in  ${\cal K}_1$ all variables occur below some function symbol 
in $\Sigma_1$, and all clauses in ${\cal K}_1$ are 
$\Sigma_1$-flat. Then, by Lemma~\ref{lemma:ext-struct}, 
we can transform $B$ into a weak partial model $B'$ of 
${\cal T}_1$ (with the $\Sigma_2$-structure inherited from $B$ and the 
$\Sigma_1$-structure inherited from $P$).
But then $B'$ weakly embeds into a total model $C$ of ${\cal T}_1$ such that 
$B'_{| \Pi_0}$ and $C_{|\Pi_0}$ are $\Pi_0$-isomorphic. 
We can use this isomorphism to transfer the (total) $\Sigma_2$-structure 
from $B$ to $C$. This way, we obtain a  
total model $A$ of ${\cal T}_0 \cup {\cal K}_1 \cup {\cal K}_2$ 
in which $P$ weakly embeds. 
The last claim is an immediate consequence of Theorem~\ref{rel-loc-embedding}.
\QED

\begin{ex}
{\em The following theory extensions satisfy  
$({\sf Emb_w})$, hence are local: 
\begin{enumerate}
\item[(1)] ${\cal E}q \subseteq {\sf Free}(\Sigma_1) \cup {\cal L}$, 
where ${\cal E}q$ is the pure theory of equality, 
without function symbols, and ${\cal L}$ the theory of lattices.
\item[(2)] ${\cal T}_0 \subseteq ({\cal T}_0 \cup {\sf Free}(\Sigma_1)) \cup 
({\cal T}_0 \cup {\sf Mon}(\Sigma_2))$, where 
$\Sigma_1 \cap \Sigma_2 = \emptyset$, and ${\cal T}_0$ is, e.g.\, 
the theory of posets.
\end{enumerate}
}
\end{ex}
An analogon of Theorem~\ref{combinations-satisfying-comp-non-disj} 
holds also in this case.

\subsection{Combinations of theory extensions satisfying $({\sf Emb_w})$}

We identify conditions under which embeddability conditions for the 
component theories imply embeddability conditions for the theory 
combination.  
\begin{thm}
\label{combine-disjoint}
Let ${\cal T}_0$ be an arbitrary theory 
in signature $\Pi_0 = (\Sigma_0, {\sf Pred})$. 
Let ${\cal K}_1$ and ${\cal K}_2$ be two sets of clauses 
over signatures $\Pi_i = (\Sigma_0 \cup \Sigma_i, {\sf Pred})$, 
where $\Sigma_1$ and $\Sigma_2$ are disjoint. 
We make the following assumptions:
\begin{quote}
\begin{itemize}
\item[(A1)] The class of models of 
${\cal T}_0$ is closed under direct limits of 
diagrams in which all maps are embeddings 
(or, equivalently, ${\cal T}_0$ is a $\forall \exists$ theory). 
\item[(A2)] ${\cal K}_i$ is $\Sigma_i$-flat and $\Sigma_i$-linear 
for $i = 1,2$, and  
${\cal T}_0 \subseteq {\cal T}_0 \cup {\cal K}_i$, $i = 1, 2$ are both local 
extensions of ${\cal T}_0$.

\item[(A3)] 
For all clauses in ${\cal K}_1$ and ${\cal K}_2$, every variable 
occurs below some extension function.
\end{itemize}
\end{quote}
Then ${\cal T}_0 \cup {\cal K}_1 \cup {\cal K}_2$ 
is a local extension of ${\cal T}_0$. 
\end{thm}
\Proof 
The proof uses the semantical characterization of locality 
in Theorems~\ref{locality-implies-embedding} 
and~\ref{rel-loc-embedding}. 
Assumption (A2) guarantees that 
the extensions ${\cal T}_0 \subseteq {\cal T}_0 \cup {\cal K}_i$, $i = 1, 2$ 
are both local  and that, 
by Theorem~\ref{locality-implies-embedding}, they satisfy 
condition ${\sf (Emb_w)}$. We show that 
${\cal T}_0 \subseteq {\cal T}_0 \cup {\cal K}_1 \cup {\cal K}_2$
satisfies condition ${\sf (Emb_w)}$, hence, 
by Theorem~\ref{rel-loc-embedding}, is local.

\medskip
Let $\Pi = (\Sigma_0 \cup \Sigma_1 \cup \Sigma_2, {\sf Pred})$ and 
let $P$ be a partial $\Pi$-algebra 
which weakly satisfies ${\cal K}_1 \cup {\cal K}_2$ and whose 
$\Pi_0$-reduct is a total model of ${\cal T}_0$. 
By the locality of the extension 
${\cal T}_0 \subseteq {\cal T}_0 \cup {\cal K}_1$, 
there exists a total  $\Pi_1$-model of 
${\cal T}_0 \cup {\cal K}_1$, 
which we denote $P_1^1$, and a weak embedding 
$\pi_1^1 : P \hookrightarrow P_1^1$.
By Lemma~\ref{lemma:ext-struct} and
Corollary~\ref{cor:ext-struct},  
a partial $\Sigma_2$-structure can be 
defined on $P_1^1$  such that $P_1^1$  
weakly satisfies ${\cal K}_2$ and $\pi_1^1$ is a weak 
$\Pi$-embedding. 

Thus, $P_1^1$ becomes a partial 
$\Pi_2$-algebra which weakly satisfies 
${\cal K}_2$, and is a total $\Pi_0$-model of ${\cal T}_0$. 
By the locality of the extension 
${\cal T}_0 \subseteq {\cal T}_0 \cup {\cal K}_2$, 
there exists a total  $\Pi_2$-model of 
${\cal T}_0 \cup {\cal K}_2$, 
which we denote $P_2^1$, and a weak embedding 
$\pi_2^1 : P_1^1 \hookrightarrow P_2^1$.
Again,  a partial $\Sigma_1$-structure can be 
defined on $P_2^1$  such that $P_2^1$  
weakly satisfies ${\cal K}_1$ and $\pi_2^1$ is a weak 
$\Pi$-embedding. 

By iterating this process we obtain a sequence of partial 
$\Pi$-structures $P_1^i, P_2^i$, 
$i \geq 1$, all of whose reducts to $\Pi_0$ are total 
models of ${\cal T}_0$, which 
weakly satisfy ${\cal K}_1 \cup {\cal K}_2$, and have the property that,  
for every $i \geq 1$, $P_1^i$ is a total $\Sigma_1$-algebra, 
$P_2^i$ is a total $\Sigma_2$-algebra, and there are 
weak $\Pi$-embeddings 
$\pi_1^i : P_1^i \rightarrow P_2^i$ and 
$\pi_2^i : P_2^i \rightarrow P_1^{i+1}$. 
$$
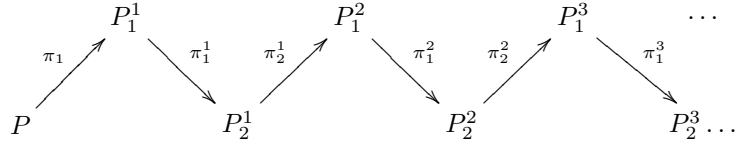
\diagram 
&P_1^1 \drto^{\pi_1^1}& &P_1^2 \drto^{\pi_1^2}& &P_1^3 \drto^{\pi_1^3}& \dots \\
P \urto^{\pi_1} & &P_2^1 \urto^{\pi_2^1}& &P_2^2 \urto^{\pi_2^2}& &P_2^3 \dots \\
\enddiagram$$
If $P_l^i$ precedes $P_k^j$ in the chain above (where $k,l \in \{1, 2\}$ and 
$i,j \geq 1$), let 
$g^{kj}_{li} : P_l^i \rightarrow P_k^j$ be the composition of the 
corresponding weak embeddings from $P_l^i$ to $P_k^j$. 
Being a composition of weak embeddings, 
$g^{kj}_{li}$ is itself a weak embedding. 

\medskip
\noindent 
Let $P ~{\scriptsize \coprod}~ (\coprod_{i \geq 1} (P_1^i \coprod P_2^i))$ be the disjoint 
union of all partial $\Pi$-structures constructed this way. 
In this disjoint union
we identify all elements that are images of the same element in 
some $P_k^i$. This is, we define an equivalence relation $\equiv$ on 
this disjoint union by 
$x \equiv y$ if $x \in P_l^i$, $y \in P_k^j$ and 
either 
(i) $P_l^i$ precedes $P_k^j$ in the chain above and $g^{kj}_{li}(x) = y$, 
or (ii) $P_k^j$ precedes $P_l^i$ in the chain above and $g^{li}_{kj}(y) = x$.
As for every 
$l \in \{1, 2\}, i \geq 1$, $g^{li}_{li}$ is the identity map,
if $x \equiv y$ for $x, y \in P_l^i$ then 
$x = y$. It is easy to see that $\equiv$ is an equivalence relation.

\medskip
\noindent 
Let $A_0 := P ~{\scriptsize \coprod}~ (\coprod_{i \geq 1} (P_1^i \coprod P_2^i))/{\equiv}$.  
We show that total functions in $\Sigma_0 \cup \Sigma_1 \cup \Sigma_2$ 
and predicates in ${\sf Pred}$ can be defined on
$A_0$ such that the expansion $A$ of $A_0$ obtained this way 
is a (total) model of 
${\cal T}_0 \cup {\cal K}_1 \cup {\cal K}_2$, 
and that the map $g : P \rightarrow A$ defined by 
$g(p) = [p]$ (the equivalence class of $p$ in $A$) is 
a weak $\Pi$-embedding. 

\medskip
\noindent 
A $\Pi$-structure on $A$ can be defined as follows:

\smallskip
\noindent
{\em Interpretation of signature $\Pi_0$.} 
We first define the $\Sigma_0$-functions.
Let $f \in \Sigma_0$ with arity $n$, and let  $[a_1], \dots, [a_n] \in A$.
Then, for every $1 \leq j \leq n$, there exist $i_j \geq 1$ such that 
$a_j \in P_1^{i_j} \coprod P_2^{i_j}$.   
Let $m = \max \{ i_j \mid 1 \leq j \leq n \}$. 
Let $b_1, \dots, b_n$ be the images of $a_1, \dots, a_n$ in $P_1^{m+1}$. 
By the definition of $\equiv$, $[b_j] = [a_j]$ for every $1 \leq j \leq n$.
$P_1^{m+1}$ is a total $\Sigma_0$-algebra, so  
$b = f_{P_1^{m+1}}(b_1, \dots, b_n)$ exists  in $P_1^{m+1}$. 
The fact that the definition does not depend on the representatives follows 
from the fact that all embeddings in the diagram are $\Sigma_0$-homomorphisms.

\smallskip
\noindent The predicates in ${\sf Pred}$ are defined in a similar way.
The fact that the definitions do not depend on the choice of 
representatives in the equivalence classes follows from the fact that all the 
maps in the diagram are $\Pi_0$-embeddings.

\medskip 
\noindent 
{\em Interpretation of the signature $\Sigma_1 \cup \Sigma_2$.} 
We define the $\Sigma_1$-functions 
(the $\Sigma_2$-functions can be defined similarly). 
Let $f \in \Sigma_1$ with arity $n$, and let $[a_1], \dots, [a_n] \in A$. 
Then, for every $1 \leq j \leq n$, there exist $i_j \geq 1$ such that 
$a_j \in P_1^{i_j} \coprod P_2^{i_j}$.   
Let $m = \max \{ i_j \mid 1 \leq j \leq n \}$. 
Let $b_1, \dots, b_n$ be the images of $a_1, \dots, a_n$ in $P_1^{m+1}$. 
By the definition of $\equiv$, $[b_j] = [a_j]$ for every $1 \leq j \leq n$.
$P_1^{m+1}$ is a total $\Sigma_1$-algebra, so  
$b = f_{P_1^{m+1}}(b_1, \dots, b_n)$ exists  in $P_1^{m+1}$. 
The equivalence class of $b$ does not depend on the choice of representatives
of the equivalence classes $[a_1], \dots, [a_n]$. Indeed, 
assume that $c_1, \dots, c_n$ are images of $a_1, \dots, a_n$ in $P_1^{k+1}$, 
with e.g.\ $k \geq m$. 
By the definition of 
$g^{1,k+1}_{1,m+1} : P_1^{m+1} \rightarrow P_1^{k+1}$,
$c_j = g^{1,k+1}_{1,m+1}(b_j)$. 
As $f_{P_1^{m+1}}(b_1, \dots, b_n)$ is defined in $P_1^{m+1}$, 
we know that $g^{1,k+1}_{1,m+1}(f_{P_1^{m+1}}(b_1, \dots, b_n))$ $=$ 
$f_{P_1^{k+1}}(g^{1,k+1}_{1,m+1}(b_1), \dots, g^{1,k+1}_{1,m+1}(b_n))$ $=$ 
$f_{P_1^{k+1}}(c_1, \dots, c_n)$. 
It follows therefore that $b \equiv f_{P_1^{k+1}}(c_1, \dots, c_n)$, 
so the equivalence class of $b$ does not depend on the choice of the 
representatives of $[a_1], \dots, [a_n]$. 
We can define $f_A([a_1], \dots, [a_n]) := [b]$. 
$f_A$ is well-defined for every $f \in \Sigma_1$.

\medskip
\noindent 
We now prove that for every $k,i$, the map 
$g_k^i : P_k^i \rightarrow A$ defined by $g(x) := [x]$ is a 
weak $\Pi$-embedding.

\smallskip
\noindent 
{\em The fact that $g_k^i$ is a $\Sigma_0$-homomorphism} is obvious.

\smallskip
\noindent 
{\em We show that $g_k^i$ is a weak $\Sigma_1$-homomorphism.}
 Let $f \in \Sigma_1$ of arity $n$ and $x_1, \dots, x_n \in P_k^i$ be 
such that $f_{P_k^i}(x_1, \dots, x_n)$ is defined. 
Then, by the definition of $f_A$, 
$f_A([x_1], \dots, [x_n]) = [f_{P_k^i}(x_1, \dots, x_n)] = 
g_k^i(f_{P_k^i}(x_1, \dots, x_n))$.  

\smallskip
\noindent
{\em The fact that $g_k^i$ is a 
$\Sigma_2$-homomorphism} can be proved analogously. 

\smallskip
\noindent {\em We prove that $g_k^i$ is injective.} 
Assume that $g_k^i(x) = g_k^i(y)$ for $x, y \in P_k^i$. 
Then $x \equiv y$, hence $g_{ki}^{ki}(x) = y$, i.e.\ $x = y$ 
(since $g_{ki}^{ki}$ is the identity map). 
This also shows that $g : P \rightarrow A$, $g(p) = [p]$ is an injective 
weak homomorphism. 

\smallskip
\noindent 
{\em We prove that $g_k^i$ is an embedding w.r.t.\ ${\sf Pred}$.}
Let $Q \in {\sf Pred}$ be an $n$-ary predicate symbol, and let 
$x_1, \dots, x_n \in P_k^i$. We show that 
$Q_{P_k^i}(x_1, \dots, x_n)$ if and only if 
$Q_A(g_k^i(x_1), \dots, g_k^i(x_n))$. By the way $Q_A$ is constructed 
it is obvious that if $Q_{P_k^i}(x_1, \dots, x_n)$ then 
$Q_A([x_1], \dots, [x_n])$. Conversely, assume that 
 $Q_A([x_1], \dots, [x_n])$. By definition, there exists $m$ and  
$b_1, \dots, b_n \in P^{m+1}_1$ such that $[x_1] = [b_1], \dots, [x_n] = [b_n]$
 and  $Q_{P^{m+1}_1}(b_1, \dots, b_n)$. The conclusion now follows from 
the fact that the composition of all maps in the diagram leading from 
$P_k^i$ to $P^{m+1}_1$ (or viceversa) is a weak $\Pi$-embedding, and 
hence also $Q_{P_k^i}(x_1, \dots, x_n)$. 

\medskip
\noindent 
The reduct to $\Pi_0$ of $A$ is the direct limit of a diagram of 
models of ${\cal T}_0$, in which all maps are embeddings. 
Therefore, if ${\cal T}_0$ is closed under such direct limits 
(i.e.\ it is a $\forall \exists$ theory) 
then $A$ is a model of  ${\cal T}_0$.

\medskip
\noindent
Finally, we show that $A$ satisfies all clauses in 
${\cal K}_1 \cup {\cal K}_2$. 
Let $C  \in {\cal K}_1$ (the case $C \in {\cal K}_2$ is similar). 
Let $\beta : X \rightarrow A$.
We know that every variable of $C$ 
occurs below a function symbol in 
$\Sigma_1$, and that all terms of $C$ containing a function symbol in 
$\Sigma_1$ are of the form $f(x_1, \dots, x_n)$. 
For every variable $x$ occurring in $C$, $\beta(x) = [a_x]$, where 
$a_x \in P_k^{j_x}$ for some $j_x \geq 1$.
Let $m = \max \{ j_x \mid x \text{ variable of } C \}$, and let $b_x$ 
be the image of $a_x$ in $P_1^{m+1}$ for each variable $x$ of $C$. 
Then $\beta(f(x_1, \dots, x_n))$ is 
defined in $P_1^{m+1}$ for every term of $C$ of the form $f(x_1, \dots, x_n)$. 
In fact, it is easy to see that for every term occurring in $C$, 
$\beta(t) = [b_t]$ for some $b_t \in P_1^{m+1}$. 
Let $\alpha : X \rightarrow P^{m+1}_1$ with  $\alpha(x) := b_x$ for every 
variable $x$ of $C$. It can be seen that 
$g_1^{m+1}(\alpha(t)) = \beta(t)$ for 
every subterm $t$ of $C$.
As $P_1^{m+1}$ satisfies $C$ and all terms in $C$ are defined under the 
assignment $\alpha$ it follows that there exists a literal $L$ in $C$ 
such that $(P_1^{m+1}, \alpha) \models_w L$.
We know that $g_1^{m+1} : P_1^{m+1} \hookrightarrow A$ is a weak embedding
w.r.t.\ $\Pi_1$.
It therefore preserves the truth of positive and negative $\Pi_1$-literals.  
Therefore, as $g_1^{m+1}(\alpha(t)) = \beta(t)$ for 
every term $t$ of $C$, $(A, \beta) \models L$. 
\QED

\begin{ex}
{\em The following combinations of theories (seen as extensions of 
the theory ${\cal T}_0$) satisfy condition 
$({\sf Emb_w})$: 

\begin{enumerate}
\item[(1)] The combination of the theory of lattices and the theory of 
integers with injective successor and predecessor is local 
(local extension of the theory of pure equality).
\item[(2)] ${\cal T}_0  \subseteq {\cal T}_0 \cup {\sf Mon}(\Sigma)$, 
where ${\sf Mon}(\Sigma) = \bigwedge_{f \in \Sigma} {\sf Mon}^{\sigma(f)}_f$, 
and ${\cal T}_0$ is one of the theories of posets, 
(dense) totally-ordered sets, 
(semi)lattices, distributive lattices, 
Boolean algebras, 
${\mathbb R}$. 
\end{enumerate}
}
\end{ex}

\section{Hierarchical and modular reasoning}
\label{hierarchic}

In what follows we discuss some issues related to modular reasoning 
in combinations of local theory extensions. By results in 
\cite{Sofronie-cade-05}, hierarchical reasoning is always possible
in local theory extensions. 
In this section we analyze possibilities of modular reasoning, 
and, in particular, the form of information which needs to be exchanged 
between provers for the component theories when reasoning in combinations of 
local theory extensions.

\subsection{Hierarchical reasoning in local theory extensions}

Consider a local theory extension 
${\cal T}_0 \subseteq {\cal T}_0 \cup {\cal K}$, where ${\cal K}$ 
is a set of clauses in the signature 
$\Pi = (\Sigma_0 \cup \Sigma_1, {\sf Pred})$.   
The locality condition
requires that, for every set $G$ of ground clauses, 
${\cal T}_1 \cup G$ is satisfiable 
 if and only if 
${\cal T}_0 \cup {\cal K}[G] \cup G$ has a weak 
partial model with additional properties.  
All clauses in ${\cal K}[G] \cup G$ have the property that 
the function symbols in $\Sigma_1$ only occur at 
the root of ground terms. 
Therefore, ${\cal K}[G] \cup G$ 
can be flattened 
and purified (i.e.\ the function symbols in $\Sigma_1$ are separated from 
the other symbols)
by introducing, in a bottom-up manner, new  constants $c_t$ for
subterms $t = f(g_1, \dots, g_n)$ with $f \in \Sigma_1$, $g_i$ ground 
$\Sigma_0 \cup \Sigma_c$-terms (where $\Sigma_c$ is a set of constants 
which contains the constants introduced by flattening, resp.\ purification), 
together with corresponding definitions $c_t \approx t$.
The set of clauses thus obtained 
has the form ${\cal K}_0 \cup G_0 \cup D$, 
where $D$ is a set of ground unit clauses of the form 
$f(g_1, \dots, g_n) {\approx} c$, where $f \in \Sigma_1$, $c$ is a 
constant,
$g_1, \dots, g_n$ are ground 
terms without function symbols in $\Sigma_1$,
and 
${\cal K}_0$ and $G_0$ are clauses without function 
symbols in $\Sigma_1$.  
These flattening and purification transformations
preserve both satisfiability and unsatisfiability with respect to 
total algebras, and also with respect to 
partial algebras in which all ground subterms 
which are flattened are defined \cite{Sofronie-cade-05}.

\medskip
\noindent
For the sake of simplicity in what follows we will always 
flatten and then purify ${\cal K}[G] \cup G$. Thus we 
ensure that  $D$ consists of ground unit clauses of the form 
$f(c_1, \dots, c_n) {\approx} c$, where $f \in \Sigma_1$, and $c_1, \dots, 
c_n, c$ are 
constants.

\begin{lem}[\cite{Sofronie-cade-05}]
Let ${\cal K}$ be a set of clauses and $G$ a set of ground clauses, 
and let ${\cal K}_0 \cup G_0 \cup D$ 
be obtained from ${\cal K}[G] \cup G$ by flattening and purification, 
as explained above. 
Assume that ${\cal T}_0 \subseteq {\cal T}_0 \cup {\cal K}$ is a local 
theory extension. 
Then the following are equivalent:

\begin{itemize}
\item[(1)] ${\cal T}_0 \cup {\cal K}[G] \cup G$ has a partial 
model in which all terms in ${\sf st}({\cal K}, G)$ are defined.
\item[(2)] ${\cal T}_0 {\cup} {\cal K}_0 {\cup} G_0 {\cup} D$ has a partial 
model with all terms in ${\sf st}({\cal K}_0,G_0,D)$ defined. 
\item[(3)] ${\cal T}_0 \cup {\cal K}_0 \cup G_0 \cup N_0$ 
has a (total) model, where 
$$ N_0  = \{ \bigwedge_{i = 1}^n c_i \approx d_i \rightarrow c = d \mid 
f(c_1, \dots, c_n) \approx c, f(d_1, \dots, d_n)\approx d \in D \}.$$ 
\end{itemize}
\vspace{-6mm}
\label{lemma-rel-transl}
\end{lem}

\subsection{Modular reasoning in local combinations of theory extensions}

Let ${\cal T}_1$ and ${\cal T}_2$ be theories 
with signatures $\Pi_1 = (\Sigma_1, {\sf Pred})$ and 
$\Pi_2 = (\Sigma_2, {\sf Pred})$, and $G$ a set of ground clauses 
in the joint signature with additional constants 
$\Pi^c = (\Sigma_0 {\cup} \Sigma_1 {\cup} \Sigma_2 {\cup} \Sigma_c, 
{\sf Pred})$.
We want to decide whether 
${\cal T}_1 \cup {\cal T}_2 \cup G \models \perp$.

\medskip
\noindent 
The set $G$ of ground clauses 
can be flattened and purified as explained above.
For the sake of simplicity, everywhere in what follows we will assume 
w.l.o.g.\ 
that $G = G_1 \wedge G_2$, where $G_1, G_2$ are 
flat and linear sets of clauses in the signatures $\Pi_1, \Pi_2$ respectively, 
i.e. for $i = 1, 2$,  $G_i = G^0_i \wedge G_0 \wedge D_i$, where 
$G^0_i$ and $G_0$ are clauses in the base theory and 
$D_i$ a conjunction of unit clauses of the form 
$f(c_1, \dots, c_n) = c, f \in \Sigma_i$. 
\begin{cor}
Assume that 
${\cal T}_1 = {\cal T}_0 \cup {\cal K}_1$ and 
${\cal T}_2 = {\cal T}_0 \cup {\cal K}_2$ 
are local extensions of a theory ${\cal T}_0$ with signature  
$\Pi_0 = (\Sigma_0, {\sf Pred})$, where $\Sigma_0 = \Sigma_1 \cap \Sigma_2$, 
and that the extension ${\cal T}_0 \subseteq {\cal T}_0 
\cup {\cal K}_1 \cup {\cal K}_2$ is local. Let $G = G_1 \wedge G_2$ be a set 
of flat, linear are purified ground clauses, 
such that $G_i = G^0_i \wedge G_0 \wedge D_i$ are as explained above.
Then the following are equivalent:
\begin{enumerate}
\item[(1)] ${\cal T}_1 \cup {\cal T}_2 \cup (G_1 \wedge G_2) \models \perp$,
\item[(2)] ${\cal T}_0 \cup ({\cal K}_1 \cup {\cal K}_2)[G_1 \wedge G_2] \cup  (G^0_1 \wedge G_0 \wedge D_1) \wedge (G^0_2 \wedge G_0 \wedge D_2) \models \perp$,
\item[(3)] ${\cal T}_0 \cup {\cal K}_1[G_1] \cup {\cal K}_2[G_2] \cup 
 (G^0_1 \wedge G_0 \wedge D_1) \wedge (G^0_2 \wedge G_0 \wedge D_2) 
\models \perp$, 
\item[(4)] ${\cal T}_0 \cup {\cal K}^0_1 \cup {\cal K}^0_2 \cup (G^0_1 \cup G_0) \cup (G^0_2 \cup G_0) \cup N_1 \cup N_2 
\models \perp$,  
where 
\begin{eqnarray*}
N_1 & = & \{ \bigwedge_{i = 1}^n c_i \approx d_i \rightarrow c = d \mid 
f(c_1, \dots, c_n) \approx c, f(d_1, \dots, d_n) \approx d \in D_1 \} \\ 
N_2 & = & \{ \bigwedge_{i = 1}^n c_i \approx d_i \rightarrow c = d \mid 
f(c_1, \dots, c_n) \approx c, f(d_1, \dots, d_n) \approx d \in D_2 \} \\ 
\end{eqnarray*}
\end{enumerate}
and ${\cal K}^0_i$ is the formula 
obtained from ${\cal K}_i[G_i]$ after purification and flattening, taking into 
account the definitions from $D_i$. 
\label{corollary-1}
\end{cor}
\Proof Direct consequence of Lemma~\ref{lemma-rel-transl}.
The fact that $({\cal K}_1 \cup {\cal K}_2)[G_1 \wedge G_2] = 
{\cal K}_1[G_1] \cup {\cal K}_2[G_2]$ is a 
consequence of the fact that $G_i$ are flattened and 
for $i = 1, 2$, ${\cal K}_i$ contains only function symbols in $\Sigma_i$. 
The equivalence of (3) and (4) follows from the fact that $\Sigma_1$ and 
$\Sigma_2$ only have function symbols in $\Sigma_0$ in common. \QED

\medskip
\noindent 
The method for hierarchic reasoning described in Corollary~\ref{corollary-1}
is modular, in the sense that once the information about  
$\Sigma_1 \cup \Sigma_2$-functions was separated
into a $\Sigma_1$-part and a $\Sigma_2$-part, 
it does not need to be recombined again. For reasoning in the 
combined theory one can proceed as follows: 
\begin{itemize}
\item Purify (and flatten) the goal $G$, and thus transform it into an 
equisatisfiable conjunction $G_1 \wedge G_2$, 
where $G_i$ consists of clauses in the signature $\Pi_i$, for $i = 1, 2$, 
and $G_i = G^0_i \wedge G_0 \wedge D_i$, as above.
\item The formulae containing extension functions in the signature $\Sigma_i$, 
${\cal K}_i[G_i] \wedge G_i$ are ``reduced'' (using the 
equivalence of (3) and (6)) to the formula 
${\cal K}^0_i \wedge G^0_i \wedge G_0 \wedge N_i$ in the base theory. 
\item The conjunction of all the formulae obtained this way, for all 
component theories, is used as input for a decision procedure for the 
base theory.  
\end{itemize}
\begin{rem}
Let  ${\cal T}_0 \subseteq {\cal T}_0 \cup {\cal K}_i$ be local 
extensions for $i = 1, 2$. Assume that ${\cal K}_i$ are $\Sigma_i$-flat and 
$\Sigma_i$-linear and 
all variables in clauses in ${\cal K}_i$ occur below a $\Sigma_i$-symbol, 
and that the extension ${\cal T}_0 \subseteq {\cal T}_0 \cup {\cal K}_1 \cup {\cal K}_2$ is local. Let $G = G_1 \wedge G_2$ be as constructed before. 
Assume that ${\cal T}_0 \cup ({\cal K}_1 \wedge G_1) \wedge 
({\cal K}_2 \wedge G_2) \models \perp$. 
Then we can construct a ground formula $I$ which contains only 
function symbols in $\Sigma_0 = \Sigma_1 \cap \Sigma_2$ such that 
\begin{eqnarray*}
({\cal T}_0 \cup {\cal K}_1) \wedge G_1  \models   I & \quad \quad & 
({\cal T}_0 \cup {\cal K}_2) \wedge G_2 \wedge I \models  \perp 
\end{eqnarray*}
\label{interpolation}
\end{rem} 
\Proof We assumed that the goal is flat and linear, i.e.\ 
$G_i = G^0_i \wedge \wedge G_0 \wedge D_i$ where $G^0_i, G_0$ 
contains only function symbols in 
$\Sigma_0$ and $D_i$ is a set of definitions of the form 
$c \approx f(c_1, \dots, c_n)$ with $f \in \Sigma_i$. If   
${\cal T}_0 \cup ({\cal K}_1 \wedge G_1) \wedge 
({\cal K}_2 \wedge G_2) \models \perp$ then, 
by Corollary~\ref{corollary-1} (with the notations used there): 
\begin{quote}
${\cal T}_0 \cup {\cal K}^0_1 \cup {\cal K}^0_2 \cup (G^0_1 \cup G_0) \cup (G^0_2 \cup G_0) \cup N_1 \cup N_2 \models \perp$.
\end{quote}
Obviously, 
every model of ${\cal T}_0$ which satisfies 
${\cal K}_1 \wedge G^0_1 \wedge G_0 \wedge D_1$ is also a model of  
${\cal T}_0 \cup {\cal K}^0_1 \cup G^0_1 \cup G_0 \cup N_1$, 
and every model of ${\cal T}_0$ 
which satisfies    
${\cal K}_2 \wedge G^0_2 \wedge G_0 \wedge D_2$ is also a model of 
${\cal T}_0 \cup {\cal K}^0_2 \cup G^0_2 \wedge G_0 \cup N_2$.
Let $I = {\cal K}^0_1 \cup G^0_1 \cup G_0 \cup N_1$. Then  
\begin{eqnarray*}
{\cal T}_1 \wedge G^0_1 \wedge G_0 \wedge D_1 & \models & I,  \\
I \wedge {\cal T}_2  \wedge G^0_2 \wedge G_0 \wedge D_2 & \models & 
{\cal T}_0 \cup ({\cal K}^0_1 \cup G^0_1 \cup G_0 \cup N_1) \cup 
({\cal K}^0_2 \cup G^0_2 \cup G_0 \cup N_2) \models \perp.
\end{eqnarray*}
All variables in clauses in ${\cal K}_i$ occur below a $\Sigma_i$-symbol,  
so ${\cal K}_i[G_i]$ (hence also ${\cal K}^0_i$) is ground for $i = 1, 2$, 
i.e.\ $I$ is quantifier-free. \QED 


\medskip  
\noindent 
If the goal is not flattened, then we can flatten and purify
it first and use Theorem~\ref{interpolation} 
to construct an interpolant $I_1$. 
We can now construct $I$ from $I_1$ by replacing each constant $c_t$ 
introduced in the purification process (and therefore contained in a
definition $c_t \approx t$ in $D_1 \cup D_2$) with the term $t$. 
It is easy to see that $I$ satisfies the required conditions. 
We can, in fact prove that only information over the shared signature 
(i.e. shared functions and constants) 
is necessary. 
\begin{thm}[\cite{Sofronie-ijcar-06}]
With the notations above, 
assume that  
$G_1 {\wedge} G_2 \models_{{\mathcal T}_1 \cup {\mathcal T}_2} \perp$.
Then there exists a ground formula $I$, containing only constants shared 
by $G_1$ and $G_2$, with $G_1 \models_{{\mathcal T}_1 {\cup} {\mathcal T}_2} I$ and $I \wedge G_2 \models_{{\mathcal T}_1 {\cup} {\mathcal T}_2} \perp$.  
\end{thm}

\section{Conclusions}
\label{conclusions}

We presented criteria for recognizing situations when 
combinations of theory extensions of a base theory are again local 
extensions of the base theory. We showed, for instance, that if 
both component theories satisfy the embeddability condition $({\sf Comp_w})$, 
which guarantees that we can always embed a partial model into one with 
isomorphic support, then the combinations of the two theories 
satisfies again condition $({\sf Comp_w})$. 
The main problem which we needed to overcome when considering more general 
combinations of local theory extensions was the preservation of
truth of clauses when extending partial operations to total 
operations in a partial algebra. We identified some conditions 
which guarantee that this is the case. 
These results allow to recognize wider classes of local 
theory extensions, and open the way for studying possibilities of 
modular reasoning in such extensions.
From the point of view of modular reasoning in such combinations of 
local extensions of a base theory, it is interesting to analyze 
the exact amount of information which needs to be exchanged between 
provers for the component theories. We showed that 
if we start with a goal 
in purified form $G = G_1 \wedge G_2$, it is sufficient 
to exchange only ground formulae containing only constants and function 
symbols common to $G_1 \wedge {\cal T}_1$ and $G_2 \wedge {\cal T}_2$. 
We would like 
to understand whether there are any links between the results 
described in this paper and other methods for reasoning 
in combinations of theories over non-disjoint signatures e.g.\  
by Ghilardi  \cite{Ghilardi04-jar}.

\medskip
\noindent {\bf Acknowledgments.} 
This work was partly
  supported by the German Research
  Council (DFG) as part of the Transregional
  Collaborative Research Center ``Automatic
  Verification and Analysis of Complex
  Systems'' (SFB/TR 14 AVACS). See
  \texttt{www.avacs.org} for more information.

\end{document}